\documentclass[aps,prl, twocolumn, superscriptaddress]{revtex4}
\usepackage{graphicx}
\usepackage{amsmath, amsthm, amssymb}

\usepackage[usenames]{color}

\newcommand{\be}{\begin{equation}}
\newcommand{\ee}{\end{equation}}
\newcommand{\bea}{\begin{eqnarray}}
\newcommand{\eea}{\end{eqnarray}}

\newcommand{\la}{\langle}
\newcommand{\ra}{\rangle}

\renewcommand{\phi}{\varphi}
\renewcommand{\epsilon}{\varepsilon}

\begin{document}

\title{Nematic Valley Ordering in Quantum Hall Systems}
\author{D. A. Abanin}
\affiliation{Department of Physics, Princeton University, Princeton, New Jersey 08544, USA}\affiliation{Princeton Center for Theoretical Science, Princeton University, Princeton, New Jersey 08544, USA}
\author{S. A. Parameswaran}
\affiliation{Department of Physics, Princeton University, Princeton, New Jersey 08544, USA}
\author{S. A. Kivelson}
\affiliation{Department of Physics, Stanford University, Stanford, California 94305, USA}
\author{S. L. Sondhi}
\affiliation{Department of Physics, Princeton University, Princeton, New Jersey 08544, USA}\date{\today}

\begin{abstract}




The interplay between quantum Hall ordering and spontaneously broken ``internal'' symmetries in two-dimensional electron systems with spin or pseudospin degrees of freedom 
 gives rise to a variety of interesting phenomena, including novel phases, phase transitions, and topological excitations.
Here we develop a theory of broken-symmetry quantum Hall states, applicable to a class of multi-valley systems, where the symmetry at issue is
a point group element that combines a spatial rotation with a permutation of valley indices.
The anisotropy of the dispersion relation, generally present in such systems, favors states where all electrons reside in one of the valleys. 
In a clean system, the valley ``pseudo-spin'' ordering, or spatial nematic ordering, occurs via a finite temperature transition.
In weakly disordered systems, domains of pseudo-spin polarization are formed, which prevents macroscopic valley and nematic ordering;
however, the resulting state still asymptotically exhibits the QHE.  We discuss the transport properties in the ordered and disordered regimes, and the relation of our results to recent experiments in AlAs.


\end{abstract}
\pacs{}

\maketitle



A remarkably diverse set of phases, exhibiting the Quantum Hall (QH) effect, are observed
in sufficiently clean two-dimensional electron systems (2DES) subjected to a high magnetic field~\cite{Pinzuk95}.
Of these, a particularly interesting subset
occurs in
multi-component QH systems, where in addition to the orbital degree of freedom within a Landau level (LL),  electrons have
low energy
``internal'' degrees of freedom, such as spin or a ``pseudo-spin'' associated with a valley or layer index. 
QH states in such systems, in addition to the
topological order, characteristic of all QH states, feature broken
global spin/pseudospin symmetries~\cite{Sondhi93,Moon95} -- a phenomenon termed QH ferromagnetism (QHFM). The entangling of the charge and spin/pseudospin degrees of freedom leads to novel phenomena in QHFM states,
including charged skyrmions~\cite{Sondhi93}, 
finite temperature phase transitions~\cite{Moon95}, and Josephson-like effects~\cite{Moon95,Pinzuk95}.

\begin{figure}
\includegraphics[width=1.6in]{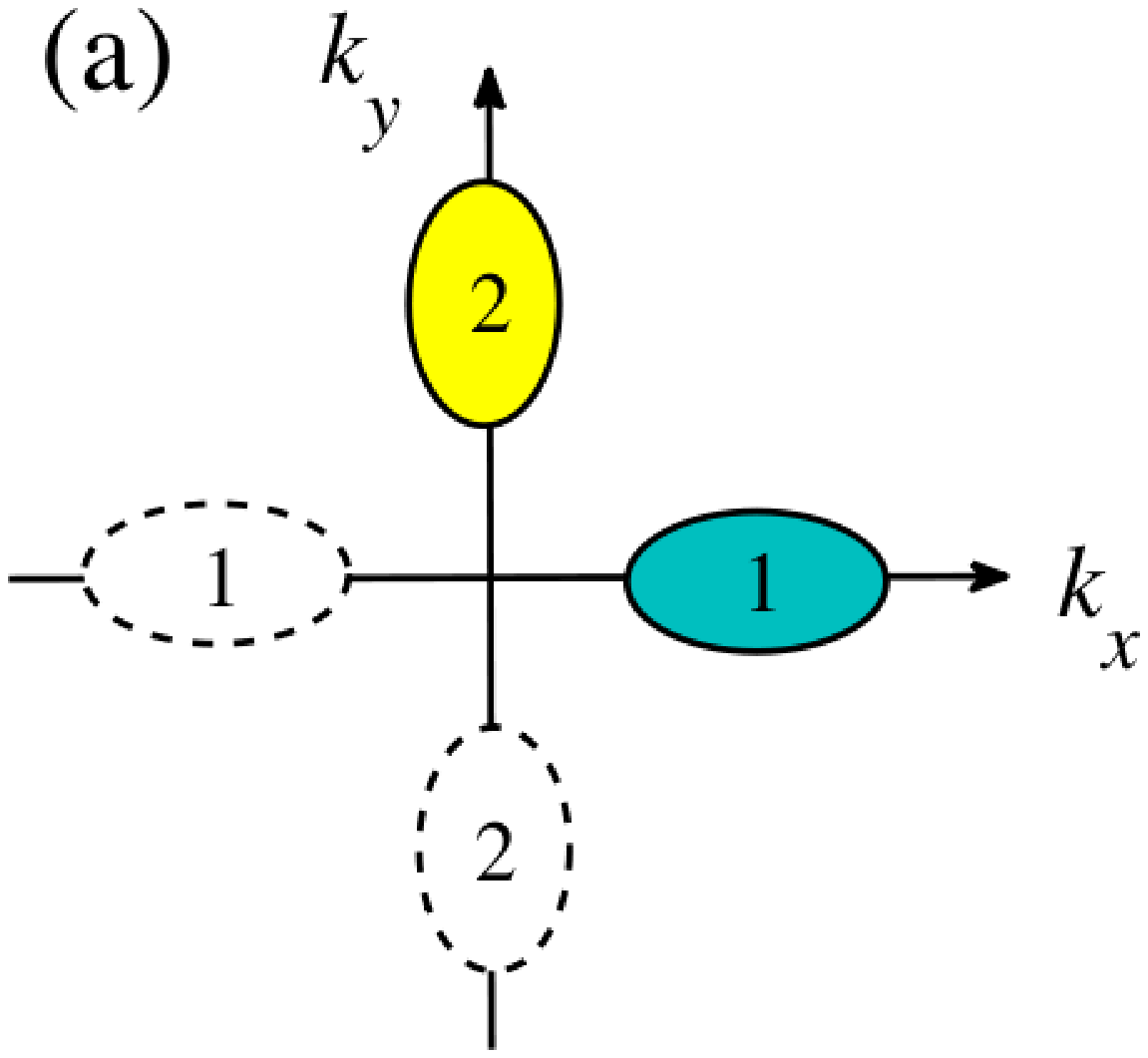}
\includegraphics[width=1.6in]{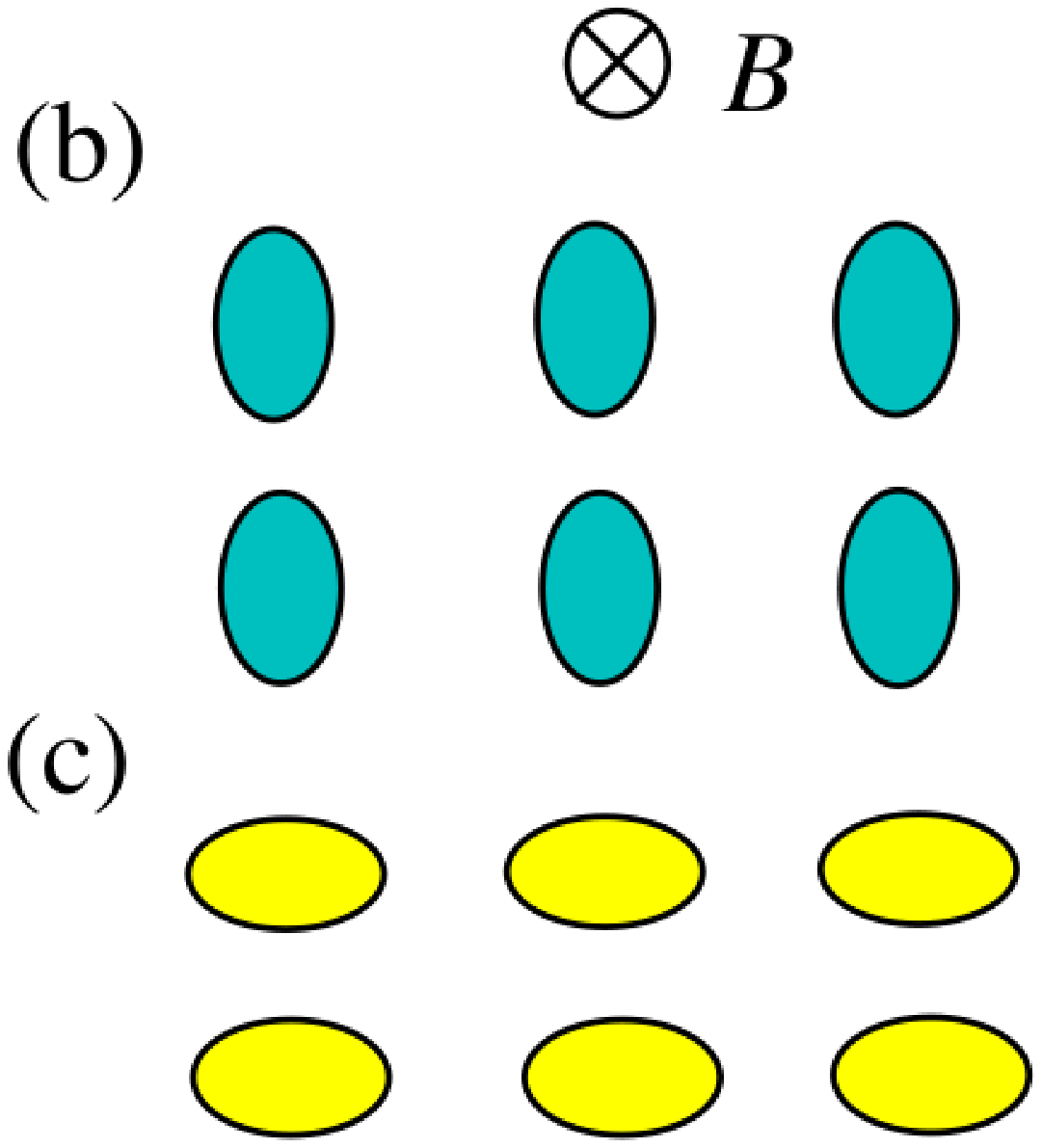}
\vspace{-1mm}
\caption[]{(a) Our model band structure. Ellipses represent lines of constant energy in the $k$-space There are two non-equivalent anisotropic valleys, $1$ and $2$. (b,c) Schematic representation of two types of order in the QHFM. The ellipses here represent LL orbitals in real space.
}
\label{fig1}
\end{figure}

In the cases studied to date, the global symmetry is an internal symmetry that acts on spin/pseudospin. In this paper we study a situation where 
the global symmetry acts simultaneously on the internal index and on the spatial degrees of freedom. This occurs naturally in a multi-valley system where different valleys are related by a discrete rotation, so that valley (pseudospin) and rotational symmetries are intertwined. An example of such a system which is central to this paper is the AlAs heterostructure~\cite{Shayegan05,Shayegan07,Shayegan09}, where two valleys with ellipsoidal Fermi surfaces are present, as illustrated in Fig.~\ref{fig1}(a).

This
linking of pseudospin and space in this system has two significant consequences at 
appropriate filling factors
such as $\nu=1$.
First, in the clean limit the onset of pseudo-spin ferromagnetism, which occurs via a
finite temperature Ising transition, is necessarily accompanied by the breaking of a rotational symmetry that corresponds to nematic order, with attendant anisotropies in physical quantities. We shall call the resulting phase a quantum Hall Ising nematic
(QHIN). (The Ising-type pseudospin ferromagnetism is consistent with the general classification of anisotropies in QHFM~\cite{Jungwirth00}).
Second, any spatial disorder, e.g. random potentials or strains, necessarily induces a random field acting on the pseudospins which thus destroys the long ranged
nematic order in the thermodynamic limit. Interestingly, though, the resulting state {\it still} exhibits the QHE at weak disorder  
so we refer to it
as the quantum Hall random field paramagnet (QHRFP).

Although for concreteness we shall focus on the simple
case of the $\nu=1$ state in AlAs-heterostructures, our findings
are readily extended to other values of $\nu$ and a variety of multi-valley systems.

{\bf Symmetries:}
The only exact symmetries
of QH systems are the discrete translational  and point group symmetries of the underlying crystalline
heterostructures.  However, in many circumstances, there are additional approximate symmetries, 
some of which are continuous.
To the extent that spin-orbit coupling can be ignored, there is an approximate U(1) spin-rotation symmetry about the direction of the magnetic field.
Since the magnetic length, $\ell_B=\sqrt{\frac{\hbar c}{eB}}$ , and the Fermi wave-length, $\lambda_F$, are long compared to the lattice constant, the effective mass approximation is always quite accurate,  so it is possible to treat the translation symmetry as continuous.
If
the electrons occupy only a valley or valleys centered on the $\Gamma$ point in the Brillouin zone, the effective mass approximation also elevates a $C_n$ point-group symmetry to a continuous U(1) rotational symmetry.  Terms which break this symmetry explicitly down to the discrete subgroup come from corrections to the effective mass approximation, and so are smaller in proportion to $(a/\lambda_F)^2$, where $a$ is the lattice constant of the semiconductor.
All three of these approximate symmetries
hold in GaAs heterostructures.

However, once there are multiple valleys centered on distinct symmetry-related Bloch wave-vectors, the effective mass tensor for each valley is, generically, anisotropic.  Thus, already in the effective mass approximation, individual valleys do not exhibit full rotational invariance;
there are only the original discrete set of rotations
which are associated with a simultaneous interchange of valleys.
 These discrete symmetries are unbroken for weak interactions in zero magnetic field. 
However, we show in this paper that
in the presence of a strong magnetic field
they are spontaneously broken at certain filling factors.

Specifically, in the two valley case considered explicitly here (Fig. 1a) the Hamiltonian has
an approximate $Z_2 \times U(1)$ invariance:  The
$Z_2$ represents the operation of a $\pi/2$ rotation combined with valley interchange.  The  $U(1)$ reflects an approximate conservation of the valley index, which is violated only by the
exponentially small Coulomb matrix elements, $V_{iv}$, which involve the intervalley scattering of a pair of electrons.
The QHFM should thus exhibit a finite-temperature $Z_2$ or Ising symmetry breaking phase transition,
accompanied by a spontaneous breaking of the rotational
 symmetry from $C_4$ to $C_2$, {\it i.e.} to Ising-nematic ordering. It is important to note that although $V_{iv}$
 breaks the approximate $U(1)$ symmetry, 
  the Ising symmetry is exact.





We should note that there is 
a
well understood counter-example to our general argument 
concerning the lack of continuous symmetries
in multi-valley systems
which is realized at a (110) surface in Si. Here the 2DEG occupies two valleys centered at
${\bf k}=\pm {\bf Q}/2$, ${\bf Q}$ being shorter than the smallest reciprocal lattice vector.  In this case, the only
 rotational symmetry is a symmetry under rotation by $\pi$.  Yet,
 to the extent the intervalley scattering, $V_{iv}$, can be neglected, this problem was shown by Rasolt {\it et al}~\cite{Rasolt86}  to have  an SU(2) pseudo-spin symmetry.
This 
derives from the fact that, in this case, the effective mass tensors in the two valleys are {\it identical}.
%
While in the case of interest to us there is only a discrete $Z_2$ symmetry, due to the effective mass anisotropy in each valley, in the limit of small anisotropy there is a
reference SU(2) symmetry which is only weakly broken. For clarity, and without loss of generality, we will
in places consider this analytically tractable limit, although in reality, the mass anisotropy in AlAs is not small.



{\bf Ising anisotropy:} 
 The single-particle Hamiltonian in each of the valleys, labeled by the index  $\kappa=1,2$, is given by
$H_{\kappa}=\sum_{i=x,y}\frac{(p_i-K_{\kappa,i}+eA_i/c)^2}{2m_{\kappa,i}}$,
where
${\bf K}_{1}=(K_0,0)$ and ${\bf K}_{2}=(0,K_0)$ are the positions of the two valleys 
  in the Brillouin zone. We
  work in Landau gauge, ${\bf A}=(0,-Bx)$, in which eigenstates can be labeled  by their momentum $p_y$ that translates into the guiding center position $X=p_y\ell_B^2$.
The lowest LL eigenfunctions in the two valleys are given by,
\be\label{eq:eigenfunction}
\psi_{\kappa,X}(x,y)=\frac{e^{ip_yy}}{\sqrt{L_y \ell_B}} \left(\frac{u_{\kappa}}{\pi} \right)^{1/4} e^{-\frac{u_\kappa(x-X)^2}{2\ell_B^2}},
\ee
where  $\lambda^2=({m_{1,x}}/{m_{1,y}})=({m_{2,y}}/{m_{2,x}})$ is the mass anisotropy in terms of which $u_1=1/u_2={\lambda}$.


Proceeding to the effects of the Coulomb interactions, we notice that the terms in the Hamiltonian that
 involve inter-valley-scattering processes require
large momentum transfer, of order $\pi/a$, and therefore they are small in proportion to $a/\ell_B$.
In accord with that, we write the Hamiltonian as follows,
\be\label{eq:coulomb}
H=H_0+H_{iv}, H_0=\frac{1}{2S}\sum_{\kappa,\kappa'} V({\bf q})\rho_{\kappa\kappa}({\bf q}) \rho_{\kappa'\kappa'}(-\bf q),
\ee
where $S=L_xL_y$ is the system's area, $\rho_{\kappa\kappa}$ is the density component within valley $\kappa$ , $V({\bf q})=\frac{2\pi e^2}{\epsilon q}$ is the matrix element of the Coulomb interaction, and $H_{iv}$ denotes inter-valley scattering terms~\footnote{Here we notice that $H_{iv}$ has a contribution $V_{iv}$ that breaks U(1) symmetry, as well as a contribution that preserves the number of electrons within each valley; the former is exponentially small in $a/\ell_B$, while the latter, although only algebraically small, has negligible value~\cite{workinprog}.}, which we neglect for now.

\begin{figure}
\includegraphics[width=3.3in]{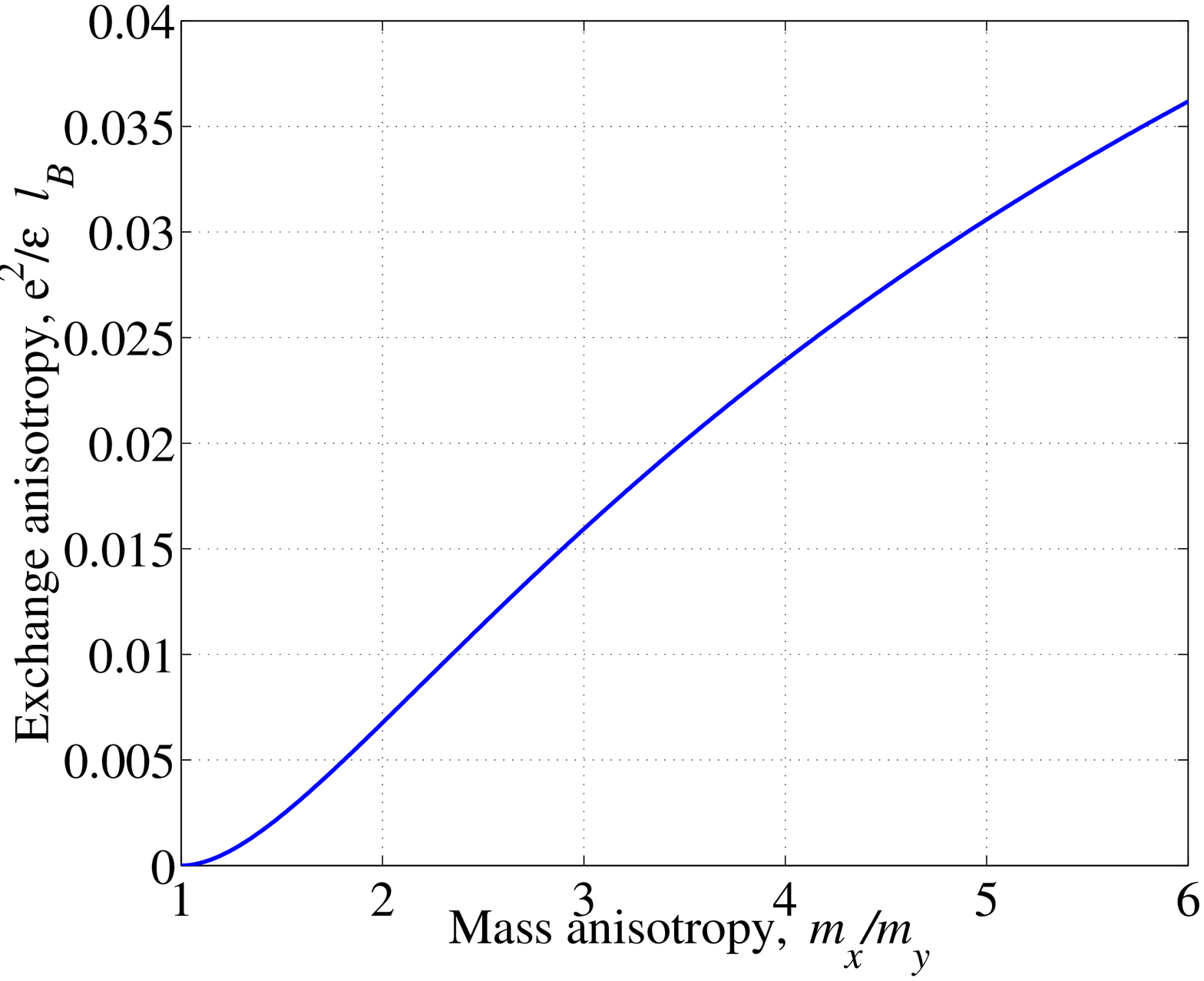}
\vspace{-1mm}
\caption[]{Easy-axis anisotropy of the QHFM as a function of underlying mass anisotropy.}
\label{fig2}
\end{figure}


To account for the spatial structure of LL wavefunctions, we follow the standard procedure of projecting the density operators onto the lowest LL (see, e.g., Ref.~\cite{Moon95}):
\be\label{eq:projected density}
\rho_{\kappa\kappa}({\bf q})=F_{\kappa\kappa}({\bf q})\bar\rho_{\kappa\kappa}({\bf q}), \,\, F_{\kappa\kappa}({\bf q})=e^{-\left(\frac{q_x^2}{4u_\kappa}+u_\kappa\frac{q_y^2}4 \right)},
\ee
where {\bf the magnetic translation operator} is given by,
$$\bar\rho_{\kappa\kappa}({\bf q})=\sum_{\bar{X}} e^{iq_x \bar{X}} c_{\kappa,X_+}^\dagger c_{\kappa,X_-}, \,\, X_{\pm}=\bar{X}\pm \frac{q_y}2. $$


In the limit of vanishing mass anisotropy, $\lambda\to 1$, the Hamiltonian $H_0$ 
is $SU(2)$-symmetric, 
so at filling factor $\nu=1$ there is a family of degenerate fully pseudo-spin polarized 
 ground states, favored by the exchange interactions.
\be\label{eq:GS}
\Psi_{\alpha,\beta}=\prod_{X} (\alpha c_{1,X}^\dagger+\beta c_{2,X}^\dagger)|0\ra,\,\, |\alpha|^2+|\beta^2|=1.
\ee
In this notation, the components of the nematic order parameter are given by $n_x=\alpha\beta^*+\alpha^*\beta, \, n_y=i\alpha\beta^*-i\alpha^*\beta, \, n_z=|\alpha|^2-|\beta|^2$, 
where ${\bf n}^2=1$.
%
We can use the states (\ref{eq:GS}) to obtain a variational estimate of the energy per electron of the system for different (uniform) values of the
order parameter  which should be reliable at least for $\lambda$ near 1.  
The result is
\be\label{eq:energy}
E_0=-\Delta_0(D_1+D_2 n_z^2), \,\, \Delta_0=\frac{1}{2}\sqrt{\frac{\pi}2} \frac{e^2}{\epsilon\ell_B}
\ee
where
\be\label{eq:Ds}
D_1=(C_1+C_2)/2, \,\, D_2=(C_1-C_2)/2,
\ee
\be\label{eq:C1}
C_1=\frac{2}{\pi}\frac{K(\sqrt{1-1/\lambda^2})}{\sqrt{\lambda}},\,
C_2=\sqrt{\frac{2\lambda}{1+\lambda^2}},
\ee
$K$ being the complete elliptic integral of the first kind. Clearly, when 
$\lambda\neq 1$, the $SU(2)$ symmetry is broken down to $Z_2\times U(1)$ and the resulting QHIN indeed has an
Ising (easy-axis) symmetry.
The magnitude of the anisotropic part of the energy, $D_2\Delta_0$, is pictured in
Fig.\ref{fig2}.
 For the experimentally relevant case, $\lambda^2\approx 5$, $\kappa\approx 10$, the anisotropy reaches a relatively large value of $5\, {\rm K}$ at 
$B=10\, {\rm T}$. Let us also note, for subsequent use, that the Ising symmetry can be
explicitly broken in experiments by the convenient application of a uniaxial strain~\cite{Shayegan05}, which then acts as a valley Zeeman field.

{\bf Thermal properties:} In order to understand the behavior of the system more generally,
and in particular to describe the properties of domain walls and excitations, we need to 
account for
spatially varying order parameter configurations. The 
classical energy functional for smooth configurations of the order
parameter 
can be obtained approximately for $|\lambda-1|\ll 1$ by  the method of Ref.~\cite{Moon95}:
\be\label{eq:energy_functional}
E[{\bf n(r)}]=\frac{\rho_s}{2}\int d^2 r (\nabla{\bf n})^2-\frac{\alpha}{2}\int d^2 r n_z^2,
\ee
where 
$\alpha\approx \frac{3}{32}\frac{\Delta_0}{2\pi\ell_B^2}(\lambda-1)^2$.
The symmetric part of the stiffness coefficient in Eq.(\ref{eq:energy_functional}) is given by,
$\rho_s=
\frac{e^2}{16\sqrt{2\pi}\epsilon \ell_B}+O(\lambda-1)$.
In writing Eq.(\ref{eq:energy_functional}), we have neglected anisotropic stiffness terms of the form,
$\frac{\rho_A}{2}\int d^2 r n_z ((\partial_x {\bf n})^2-(\partial_y {\bf n})^2)$, $\frac{\rho_{A'}}{2}\int d^2 r [3(\nabla n_z)^2-(\nabla{\bf n})^2]$.  While these terms are also quadratic in the gradient expansion, in the limit $|\lambda-1|\ll 1$, the first term is at most cubic~\cite{workinprog} in $\lambda-1$, such that $\rho_A\approx o ((\lambda-1)^2)$, while the second term is quadratic,  $\rho_{A'}=o((\lambda-1))$, and so they are much smaller than the gradient term we have kept.

The nematic ordering temperature can readily be estimated from Eq.(\ref{eq:energy_functional}), which is precisely the continuum limit of the 2D Heisenberg ferromagnet with weak Ising anisotropy.  Consequently, $T_c$ vanishes for $\alpha=0$, but only logarithmically, due to the exponential growth of correlations in the Heisenberg model,
\be
k_B T_c \sim 4\pi \rho_s \log^{-1}[\rho_s/\alpha \ell_B^2].
\ee
Since, in reality, the anisotropy is not small, a more robust estimate  is just $T_c \sim \rho_s$.  This puts it in the range of several Kelvin, well above typical temperatures at which quantum Hall experiments are carried out, which range from a few tens to a few hundred milli-Kelvins~\cite{Shayegan05}.

\noindent
{\bf Domain walls and quasiparticles:} The topological defects of an Ising ferromagnet are domain walls, in this case domain
walls across which the valley polarization  changes sign.  We obtain a domain wall solution by minimizing the classical energy
in
 Eq.(\ref{eq:energy_functional}) to obtain the length scale $L_{0} = \sqrt{\rho_s \over \alpha}$
which characterizes the domain-wall width, and the surface tension, ${\cal J}\sim \sqrt{\rho_s\alpha}$, its creation energy per unit length.  The domain wall solution obtained in this way spontaneously breaks the approximate $U(1)$ symmetry as the energy is independent of the choice of the axis of rotation of ${\bf n}$ in the plane perpendicular to $n_z$.  Naturally, since the domain wall is a one dimensional object, thermal or quantum fluctuations restore the symmetry, but at $T=0$, and in the absence of explicit symmetry breaking perturbations, what remains is a gapless ``almost Goldstone mode'' and power-law correlations along the domain wall.  A small gap in the spectrum and an exponential fall-off of correlations beyond a distance $\xi_{iv}$ are induced when the effects of the weak intervalley scattering terms, $V_{iv}$, are included.

In AlAs, the anisotropy is
$\lambda^2 \approx 5$, so $L_{0}$ is only 30\% greater than $\ell_B$, which indicates that our
treatment should be supplemented by microscopic calculations that
can better handle a strong Ising anisotropy \cite{workinprog}.
A variational ansatz for a domain-wall can be constructed of the same form as in Eq. \ref{eq:GS} by treating $\alpha$ and $\beta$ as (complex) functions of $X$, with asymptotic forms $(\alpha,\beta) \to (1,0)$ as $X\to -\infty$ and  $(\alpha,\beta) \to (0,1)$ as $X\to \infty$. In the limit of large $\lambda$, the optimal such state consists of a discontinuous jump between these two limiting values across the domain wall, so that the domain wall width is simply equal to $\ell_B$.
When inter-valley scattering is absent, such a wall, being a boundary between two different QH liquids (one with $\nu_1=0,\nu_2=1$, the other with $\nu_1=1,\nu_2=0$), supports two counter-propagating chiral gapless modes---
one with pseudo-spin ``up'' and the other with pseudo-spin ``down.''  
 Coulomb interactions between the two modes turn this into a type of Luttinger liquid.
 This connects smoothly to the description obtained above in the limit of weak anisotropy, and indeed
 the Luttinger liquid action can be derived explicitly from a $\sigma$-model description~\cite{Falko00,MitraGirvin} by augmenting the classical energy in Eq.~(\ref{eq:energy_functional}) with an appropriate quantum dynamics. 

The other excitations of interest are charged quasiparticles and it is well known that in the $SU(2)$ limit at $\lambda=1$ they
are pseudospin skyrmions of divergent size~\cite{Sondhi93}. However, the smallness of $L_{0}$ at $\lambda^2 \approx 5$ alluded to
above implies that for the experimentally relevant case the quasiparticles will be highly, if not completely, valley polarized.

{\bf Properties of the clean system:}
%
For $T<T_c$, where the pseudospin component $n_z$ 
has a nonzero
expectation value, 
$C_4$ rotation symmetry is 
spontaneously broken to $C_2$.  Thus, non-zero values of
any non-trivial traceless
symmetric tensor can also be used as an order parameter.

Ideally, thermodynamic quantities, for instance of the difference in the valley occupancies, provide the conceptually simplest measures of the broken symmetry.  However, such quantities are not easily measured in practice.
Following our remark above, we should just as well be able to use the experimentally accessible transport
anisotropy ratio
\be\label{eq:transport_anisotropy}
N=\frac{\sigma_{xx}-\sigma_{yy}}{\sigma_{xx}+\sigma_{yy}}\neq 0
\ee
as a 
measure of nematic order.
However, at 
$T=0$, where $\sigma_{aa}=0$, $N$ is ill-defined.  This problem can be resolved by 
  measuring $\sigma_{aa}$ at  finite temperature, $T >0$, and then 
   taking the limit $T \to 0$.  (Alternatively,  one could imagine working at finite frequency, and then taking the limit as the frequency tends to 0.)
However, in practice, the conductivity is strongly affected by the presence of even weak disorder, so any practical discussion of the resistive anisotropy must be preceded by an analysis of the effects of disorder.

{\bf Length scales from  weak disorder:}
By analogy with the random-field Ising model~\cite{Binder-RFIM}, we know~\cite{Aizenman89} that even an arbitrarily weak random valley-Zeeman
field destroys the ordering of the QHIN, leading to formation of ``Imry-Ma'' domains of opposite valley-polarization.
%
 In AlAs such disorder can stem from random strains, which lead to position-dependent relative shifts of the energies of the two valleys. 
 While the average strain ({\it i.e.} the average pseudo-magnetic field) can be externally controlled~\cite{Shayegan05}, fluctuations of the strain are inevitable. Random fluctuations of the electric potential, $V$, also give rise to a random valley field.

 The  coupling of random strain and potential disorder to the QHIN order parameter is 
\be\label{eq:random strain}
E_{st}=\frac{1}{2}\int d^2 r\, h({\bf r})n_z({\bf r}),
\ee
where  the random field $ h({\bf r})=[h_{st}({\bf r}) + h_{pot}({\bf r})]$ with $h_{st}({\bf r})\propto \frac{\partial u({\bf r})}{\partial x}-\frac{\partial u({\bf r})}{\partial y}$, $u({\bf r})$ being the displacement of point ${\bf r}$ of the crystal, and
%
%
\be\label{eq:random_field}
h_{pot}({\bf r})=\frac{(m_x-m_y)\ell_B^2}{2\pi\hbar^2} \left[\left(\frac{\partial V}{\partial x}\right)^2-\left(\frac{\partial V}{\partial y}\right)^2 \right].
\ee
%
%
On the basis of this analysis, we expect that the random valley-Zeeman field is smooth, with a typical correlation length $\ell_{\rm dis}\gg \ell_B$.
For weak disorder, the Imry-Ma domain size is set by the mean-squared strength of the random field (assumed to have zero mean and to be short-range correlated)
$W\equiv \int d^2r \langle h({\bf r}) h({\bf 0})\rangle$ 
and the domain wall energy per unit length, ${\cal J}$ (defined above) as
$\xi_{IM} \propto \exp[ \ A ({\cal J})^2/W\ ]$ where $A$ is a number of order 1.  Because of the exponential dependence on disorder, it is possible for $\xi_{IM}$ to vary, depending on sample details, from microscopic to macroscopic length scales.

Disorder
also leads to scattering between the valleys although this is again suppressed due to the mismatch between the reciprocal
lattice vector and the length scale of the dominant potential fluctuations.  There is thus a second emergent length scale, $\xi_{iv}$, which is the length scale beyond which conservation of valley pseudo-spin density breaks down.  However, this length scale approaches a finite value in the limit of vanishing disorder due to intervalley Coulomb scattering, discussed above.  Different regimes of physics are possible depending on the ratio of $\xi_{IM}/\xi_{iv}$.  Finally, especially when the filling factor deviates slightly from $\nu=1$, there is a length scale, $\xi_{QP}$, which characterizes the quasi-particle localization length.  Because the magnetic field quenches the quasiparticle kinetic energy, even for extremely weak disorder, we expect $\xi_{QP} \sim \ell_B$ is relatively short.

{\bf Intrinsic resistive anisotropy:}
In a quantum Hall state at low temperatures, dissipative transport is usually due to hopping of quasiparticles between localized states, accompanied by energy transfer to other degrees of freedom~\cite{EfrosShklovskii}. Typically, transport is of variable-range-hopping (VRH) type, such that the optimal hop is determined by the competition between 
the energy offset of the two states and their overlap. We will now apply these ideas to our system when its transport primarily involves hopping of electrons between localized states within one of the valleys. This requires a) either that a uniaxial strain
be applied to  substantially eliminate domain walls and achieve valley polarization in the proximity of $\nu=1$ or that
the sample be smaller than $\xi_{IM}$
and b) that $\xi_{iv}$ be large compared to $\xi_{QP}$. For each valley the localization length is anisotropic, owing to the mass anisotropy, which results in the anisotropy of the corresponding contribution to the VRH conductivity.

The contribution to the resistive anisotropy from quasiparticles in valley 1, $N_1$, can be computed as follows:
First, we transform the  anisotropic VRH problem into the isotropic one by rescaling coordinates, $x=\tilde{x}/\sqrt{\lambda}, \, y=\sqrt{\lambda}\tilde{y}$.
In the new coordinates the effective mass tensor is isotropic, which,
given the uncorrelated nature of the potential, implies that the VRH problem is isotropic~\footnote{Under the assumption that
the energy transfer in VRH is due to phonons, and the electron-phonon coupling is scalar.}, and therefore $\tilde{\sigma}_{xx}=\tilde{\sigma}_{yy}$. Since the ratio of the conductivities in the original coordinates is given by $\frac{\sigma_{xx}}{\sigma_{yy}}=\frac{1}{\lambda^2}\frac{\tilde\sigma_{xx}}{\tilde\sigma_{yy}}$, \be\label{eq:transport_anisotropy2}
N_1=
\frac{1-\lambda^2}{1+\lambda^2}, \ee
which is negative for $\lambda>1$, as expected, {\it i.e.} it is more difficult for particles to move in the direction of larger mass.
Clearly, the resistive anisotropy produced by quasiparticles  in valley 2 is $N_2=-N_1$. 

At $\nu=1$ localized states in both valleys are present, and due to combined particle-hole/valley-reversal symmetry of the state (in the absence of Landau-level mixing), the density of localized states should be same:
The resistivity is thus expected to be isotropic!
However,
%
for $\nu \neq 1$, particle hole symmetry is broken.
Consider the case in which 
 $n_z=+1$, which corresponds to filling valley $1$ states. Then, at  slightly filling factor, $\nu=1-\delta\nu$ with $1 \gg\delta\nu>0$, the density of localized states for valley $\kappa=1$
 exceeds that for valley $\kappa=2$. Due to 
 the exponential sensitivity of the VRH conductivity to the density of states, this implies that the contribution of valley 1 to the total conductivity 
 dominates, leading to an anisotropy of the total conductivity $N\approx N_1$.  Conversely, for $\nu=1+\delta \nu$, $N\approx N_2 = -N_1$!
It is worth noting that the scaling argument presented above for VRH regime is likely more general, and also applicable to the regime of thermally activated transport, which is relevant at intermediate temperatures.

{\bf Domain walls and the QHRFPM:} We now move away from the above limit to where domain walls are a significant
contributor to the transport---to systems much bigger than $\xi_{IM}$ and at weak uniaxial strain. Now,
dissipative transport is complicated by the existence of multiple emergent length scales.  Transport within a nematic domain proceeds by variable range  hopping and/or thermal activation of quasiparticles.  For length scales larger than $\xi_{IM}$, it is likely to be dominated by transport along domain walls, which will have insulating character or metallic character depending on whether viewed at distances large or small compared to $\xi_{iv}$.

A key question is whether the QHE survives the formation of domains. This is trickiest when no net valley Zeeman field
is applied where in the thermodynamic limit the domain walls form a percolating network.
In the limit $\xi_{iv}\to\infty$, the associated edge channels  are conducting, 
and the domain wall network can be expected to be well described by two copies of the Chalker-Coddington network
model~\cite{Chalker} at criticality. This implies a critical metallic longitudinal DC conductivity of order
$e^2/h$ and the absence of QHE.
However, at length scales longer than $\xi_{iv}$ (or temperatures less than $T_{iv} \sim 1/\xi_{iv}$)
the domain wall states are localized, which implies a phase that exhibits the QHE without Ising/Nematic order---the
QHRFPM.
  Needless to say, in the absence of substantial amount of short ranged disorder (which can produce a relatively small value of $\xi_{iv}$) the topological (quantum Hall) order in the QHRFP is likely to be fragile .

%
%
(In some ways similar results were obtained by Rapsch et al.~\cite{Lee03}, who considered an SU(2)-symmetric disordered QHFM, where
magnetic order is destroyed by forming a spin glass without destroying the QHE. )

In the presence of a uniform valley Zeeman field $\bar h$ the existence of the QHE is much more robust.
Even weak fields can restore a substantial degree of valley density polarization as
domains aligned with this field grow while those aligned
opposite to it shrink. 
Consequently, the domain walls no
longer percolate but rather are separated by a finite distance
that grows with increasing $\bar h$. 
While we have yet to
construct a detailed 
theory of the transport in this regime, it is clear that the characteristic energy scale characterizing the dissipative transport will rise rapidly from $T_{iv}$ for $\bar h=0$, to the clean-limit gap $\Delta_0\sim \rho_s$ for substantial values of $\bar h$.  It is also important to note that this equilibrium response will come embedded in a matrix of
dynamical phenomena characteristic of the random field Ising model that can be translated straightforwardly to the case of the Ising-nematic, as has been discussed in another context in Ref. \cite{carlson}.  In particular, the macroscopic
nematicity induced by the application of $\bar h$ will be metastable for long times, even upon setting
$\langle h \rangle = 0$---thus giving rise to hysteresis.

{\bf Experiments:} Turning briefly to experiments we note that an anomalously strong strain induced enhancement of the
apparent activation
energy at $\nu=1$ has been observed\cite{Shayegan05} in  AlAs, where it was tentatively attributed to the occurrence of valley skyrmions.
As we noted earlier, in view of our estimate of a large Ising anisotropy skyrmions of the requisite size (about 15
flipped pseudospins) are implausible.
We would like to suggest that it is more plausible that these remarkable observations are associated with the growth of QHIN domains. In support of this idea, we have estimated the domain size from the long ranged part of the potential
disorder alone and find that it should be order the distance to the dopant layer, $\xi_{\rm dis}\approx 50 \,{\rm nm}$
which is thus much smaller than the system size. However, we currently lack a plausible estimate of $\xi_{iv}$ which is sensitive to the short ranged part of the disorder and which is needed to round out this explanation.
Direct measurements of resistive anisotropies, and of hyseretic effects characteristic of the random field Ising model~\cite{carlson} could directly confirm this proposal.


\noindent
{\bf Related work:} We note that there is a sizeable body of existing work on Ising QHFMs produced at level crossings
of different orbital LLs, which is typically achieved by applying tilted magnetic fields. These systems exhibit enhanced
dissipation at coincidence~\cite{Ising00,Ising0,Ising1,Ising2} which is the analog of a
dissipation peak at zero valley Zeeman field in our language. Qualitatively, our results are consistent with this
earlier work.
Where we differ is in our contention
that the domain walls do not, even at zero valley Zeeman field, produce dissipation at $T = 0$---in the previous work~\cite{MacDonald00,Chalker02}
this was not explicitly addressed in part as the focus was
on accounting for the unexpected dissipation. The reader will also note that the QHIN studied here differs from the
``nematic quantum Hall metal'' (NQHM) phase which has been  observed~\cite{Eisenstein00}  in ultra-clean GaAs-GaAlAs heterostructures for fields at which the $n > 1^{st}$ Landau level is nearly half filled. Unlike our system, the NQHM is a metallic state which does not exhibit the QHE, but has a strongly anisotropic resistivity tensor.

\noindent
{\bf In closing:} The distinctive feature of our system is the breaking of a global symmetry that combines spatial
and internal degrees of freedom. This physics and its attendant consequences will generalize immediately to
other ferromagnetic fillings in the present system and then to other experimentally established examples of multi-valley
systems such as monolayer and bilayer graphene~\cite{Zhang06,Feldman09}, where two valleys are present, and Si
(111)~\cite{Kane07}, where, depending on the parallel field, either 4 or 6 degenerate valleys can be present. Potentially, our ideas could apply farther afield in the case of 3D Bi, where three electron pockets
related by $2\pi/3, 4\pi/3$ rotations are present. Recently, high-field anomalies in transport and thermodynamic properties of Bi were found~\cite{Li08,Behnia08}, which may indicate spontaneous breaking of the $Z_3$ valley symmetry driven by magnetic field, reminiscent of QHFM.

\noindent
{\bf Acknowledgements.} We would like to thank David Huse and Steve Simon for insightful discussions, and Mansour Shayegan, Tayfun Gokmen,  Medini Padmanabhan for helpful discussions and sharing their unpublished data. We also thank John Chalker, Allan MacDonald, Boris Shklovskii for very helpful correspondence and comments on the manuscript.

\end{document}